# FARSIDE: A Low Radio Frequency Interferometric Array on the Lunar Farside

*Electromagnetic Observations from Space*

**Principal Authors:** Jack O. Burns (University of Colorado Boulder) & Gregg Hallinan (California Institute of Technology)

**Co-authors:** Jim Lux (JPL/Caltech), Andres Romero-Wolf (JPL/Caltech), Lawrence Teitelbaum (JPL/Caltech), Tzu-Ching Chang (JPL/Caltech), Jonathon Kocz (Caltech), Judd Bowman (ASU), Robert MacDowall (NASA Goddard), Justin Kasper (University of Michigan), Richard Bradley (NRAO), Marin Anderson (Caltech), David Rapetti (University of Colorado Boulder)

**Lead author contact info:** Jack Burns, jack.burns@colorado.edu; Gregg Hallinan gh@astro.caltech.edu

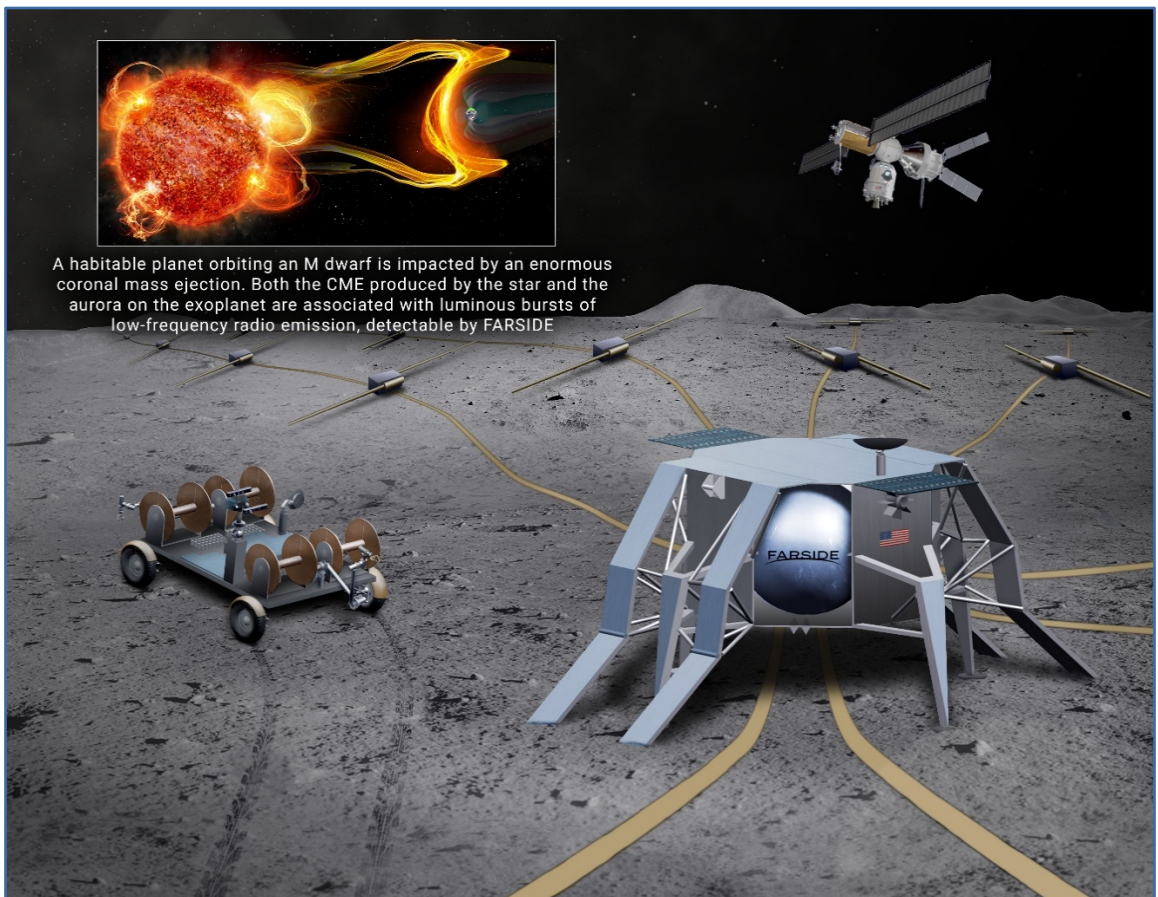

*The FARSIDE mission concept consists of a 128-element, radio interferometer, a base station that provides power, signal processing and telecommunications to the Lunar Gateway, and a rover that deploys and assembles the array. The lunar farside provides isolation from interference and is the only location within the inner solar system for sky noise-limited 200 kHz - 40 MHz radio frequency observations.*

# EXECUTIVE SUMMARY

FARSIDE (*Farside Array for Radio Science Investigations of the Dark ages and Exoplanets*) is a Probe-class concept to place a low radio frequency interferometric array on the farside of the Moon. A NASA-funded design study, focused on the instrument, a deployment rover, the lander and base station, delivered an architecture broadly consistent with the requirements for a Probe mission. This notional architecture consists of 128 dual polarization antennas deployed across a 10 km area by a rover, and tethered to a base station for central processing, power and data transmission to the Lunar Gateway. FARSIDE would provide the capability to image the entire sky each minute in 1400 channels spanning frequencies from 100 kHz to 40 MHz, extending down two orders of magnitude below bands accessible to ground-based radio astronomy. The lunar farside can simultaneously provide isolation from terrestrial radio frequency interference, auroral kilometric radiation, and plasma noise from the solar wind. It is thus the only location within the inner solar system from which sky noise limited observations can be carried out at sub-MHz frequencies. This would enable near-continuous monitoring of the nearest stellar systems in the search for the radio signatures of coronal mass ejections and energetic particle events, and would also detect the magnetospheres for the nearest candidate habitable exoplanets. Simultaneously, FARSIDE would be used to characterize similar activity in our own solar system, from the Sun to the outer planets, including the hypothetical Planet Nine. Through precision calibration via an orbiting beacon, and exquisite foreground characterization, FARSIDE would also measure the Dark Ages global 21-cm signal at redshifts z~50-100. The unique observational window offered by FARSIDE would enable an abundance of additional science ranging from sounding of the lunar subsurface to characterization of the interstellar medium in the solar system neighborhood.

# 1    KEY SCIENCE GOALS AND OBJECTIVES

## 1.1    The Magnetospheres and Space Weather Environments of Habitable Planets

The discovery of life on a planet outside our solar system is at the heart of NASA's Science Mission Directorate. Such a discovery may arrive within the next few decades and is the focus of a number of planned and concept NASA missions. The most likely avenue involves spectral observations of biosignature gases, such as $O_2$, $O_3$, $CH_4$, and $CO_2$ on an Earth-like planet orbiting a nearby star. A tiered approach would involve discovery (e.g., *TESS*, ground-based Radial Velocity [RV] surveys), characterization and eventual deep coronograph-assisted observations by missions such as *JWST*, *WFIRST*, *HabEx* and *LUVOIR* as well as ground-based extremely large telescopes.

*The Active Young Sun:* However, it is becoming increasingly apparent that both the selection of candidate exoplanets for deep searches for biosignatures, and interpretation of the observed atmospheric composition, must take into account the space weather environment of the host star, and whether the planet possesses a large-scale magnetosphere capable of retaining an atmosphere within this space environment. The enhanced radiative output at higher energies during flares leads to strong expansion of planetary atmospheres [6-8]. Simultaneously, the stellar wind of a young star is much denser and faster, compressing the magnetosphere, particularly during coronal mass ejections (CMEs), which are also presumed to be much more frequent [9][1], as well as associated solar energetic particle (SEP) events. The long-term impact of such activity was recently established in dramatic fashion by the Mars Atmosphere and Volatile EvolutioN (*MAVEN*) mission, which confirmed that ion loss due to solar CMEs early in Mars history likely severely depleted its atmosphere [10].

*Extreme M Dwarfs:* The impact of magnetic activity on planets orbiting M dwarfs in particular has become a topic of increasing significance. Many M dwarfs are known to be particularly magnetically active, flaring frequently and with much higher energy than produced in solar flares [11]. Moreover, M dwarfs have longer spin-down timescales and are thus magnetically active for a much longer period of time than solar-type stars [12]. Studies of possible flares and CME events on planets in the habitable zone around such stars suggest that these events severely impact the ability of such planets to retain their atmospheres [13, 14], particularly as the habitable zone is much closer to the parent star than the solar case.

---

[1] And all references therein.

*Detecting Stellar CMEs and SEP events:* While these modeling results paint a potentially bleak picture, there is a very large degree of uncertainty in the space environments to which exoplanets are exposed. Other than the Sun, no main sequence star has been detected to produce a CME. Similarly, detection of exoplanet magnetic fields has yet to be achieved and remains the most crucial ingredient in assessing planetary habitability in the context of stellar activity.

Solar CMEs and SEP events can be accompanied by radio bursts at low frequencies, particularly so-called Type II bursts, as well as a subset of Type III bursts (complex Type III bursts). The emission is produced at the fundamental and first harmonic of the plasma frequency and provides a diagnostic of the density and velocity (few 100 to >1000 km/s) near the shock front, while the flux density of the burst depends sensitively on the properties of the shock and solar wind [15].

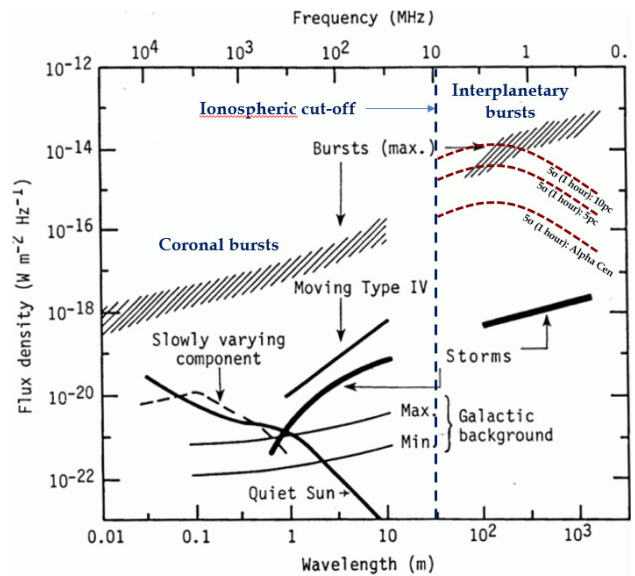

Solar radio bursts are likely the most intense sources of extraterrestrial radio emission ever observed from Earth (Figure 1), reaching a flux density $> 10^{-14}$ W m$^{-2}$ Hz$^{-1}$ ($>10^{12}$ Jy). However, the brightest bursts are rare and peak at frequencies <10 MHz, which can account for the non-detection of such bursts from solar-type stars. The non-detection of such events from active M dwarfs has been surprising [16, 17]. Even slow rotating M dwarfs produce a "superflare" ($10^{34}$ erg) each month, with >100x larger energy than any recorded on the Sun [4], and a correlation is observed between the flare energy and mass (or kinetic energy) of CMEs observed on the Sun [1, 18] which, if applicable to M dwarfs, should produce very large CMEs, and associated luminous Type II bursts.

**Figure 1:** The flux density and frequency/wavelength of the brightest radio sources observed from Earth orbit. The brightest phenomena are solar radio bursts at all frequencies, particularly Type II and III bursts [2]. Interplanetary bursts are the most luminous and are not detectable from the ground. FARSIDE can detect such events out to >10 pc.

However, this scaling law may not extend up to the flare energies observed for M dwarfs. For example, it is possible that CMEs are confined or significantly suppressed by the presence of a strong large-scale magnetic field [19]. Although the largest events would still scape to interplanetary distances, this would lessen the potential impact of CMEs/SEPs on the atmospheres of orbiting exoplanets. An alternative possibility is that CMEs are present, but are not detected, simply because the Alfvén speed is too high in the coronae of M dwarfs for a shock to form [16, 20]. The presence of a shock is an observed necessary condition for Type II emission (and a prerequisite for the generation of most SEP events). Shocks can potentially form at much greater distances from the star, where the Alfvén speed drops below the velocity of the CME, but the associated Type II burst would then be at frequencies below 10 MHz, and undetectable to ground-based radio telescopes.

FARSIDE would detect the equivalent of the brightest Type II and Type III bursts out to 10 pc at frequencies below 10 MHz. By imaging >10,000 degrees every 60 seconds, it would monitor a sample of solar-type stars simultaneously, searching for large CMEs. For the Alpha Cen system, with two solar-type stars and a late M dwarf, it would probe down to the equivalent of $10^{-15}$ W m$^{-2}$ Hz$^{-1}$ at 1 AU, a luminosity at which solar radio bursts are frequently detected [21]. The nearby young active solar-type star, Epsilon Eridani (spectral type K2), is another priority target. For the case of M dwarfs, FARSIDE would be able to detect Type II bursts formed at the distance where super-Alfvénic shocks should be possible for M dwarfs and directly investigate whether the relationship observed between solar flares and CMEs extends to M dwarfs. If it does, FARSIDE should also detect a very high rate of radio bursts from this population.

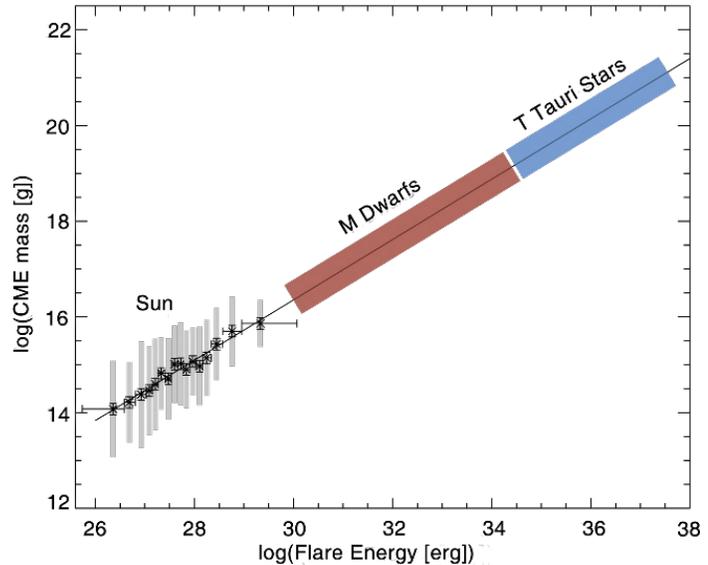

**Figure 2:** The relationship between flare energy and CME mass observed for the Sun [1], extrapolated to flare energies observed on M dwarfs [4] and young stars.

*Exoplanet Magnetospheres:* All the magnetized planets in our solar system, including Earth, produce bright coherent radio emission at low frequencies, predominantly originating in high magnetic latitudes and powered by auroral processes [22]. Detection of similar radio emission from candidate habitable planets is the only known method to detect the presence and strength of their magnetospheres. The temporal variability of this radio emission can also be used to determine the rotation periods and orbital inclinations of these planets. The radio emission, generated by the electron cyclotron maser instability, is produced at the electron cyclotron frequency, ~2.8 x $B$ MHz, where $B$ is the magnetic field strength in Gauss at the source of the emission. Notably, only Jupiter has a strong enough magnetic field to be detected from the ground. The radio emission of Earth, Saturn, Uranus and Neptune are predominantly confined to frequencies < 1 MHz requiring a space-based instrument for detection.

Extending to the exoplanet domain requires a very large collecting area at low frequencies, with the first detections likely to be ground-based. Indeed, in a recent breakthrough, radio emission has been detected from a possible free-floating planetary mass object (12.7 +/- M$_{jup}$; [23]). This is the first detection of its kind and confirmed a magnetic field much higher than expected, >200 times stronger than Jupiter's, reinforcing the need for empirical data. However, the detection of the magnetic fields of candidate habitable planets will almost certainly require a space-based array, if the magnetic fields are within an order of magnitude in strength of Earth's magnetic field. Detection of the magnetic field of planets orbiting in the habitable zone of M dwarfs is particularly key, as such planet may require a significantly stronger magnetosphere than Earth to sustain an atmosphere [14].

The detection of radio emission from planets orbiting nearby stars is very sensitive to the stellar wind conditions imposed by its parent star. During enhanced solar wind conditions, the Earth's radio emission can increase in luminosity by orders of magnitude (Figure 3). Therefore, as is the case for detecting Type II bursts associated with CMEs, it is essential to have the capability to monitor large numbers of planets simultaneously. Predictions for terrestrial exoplanets in the habitable zone of M dwarfs, and thus embedded within a dense stellar wind environment, predicts radio luminosities that are orders of magnitude higher than the Earth. Figure 3 highlights that FARSIDE would detect the radio emission from a population of rocky planets orbiting nearby M dwarfs, including a number of candidate habitable planets, providing the first measurements of terrestrial planet magnetospheres outside our solar system. Detection of such magnetospheres, if present, would identify the most promising targets for follow-up searches for biosignatures, as well as providing a framework for comparative analysis of exoplanet magnetospheres and atmospheric composition.

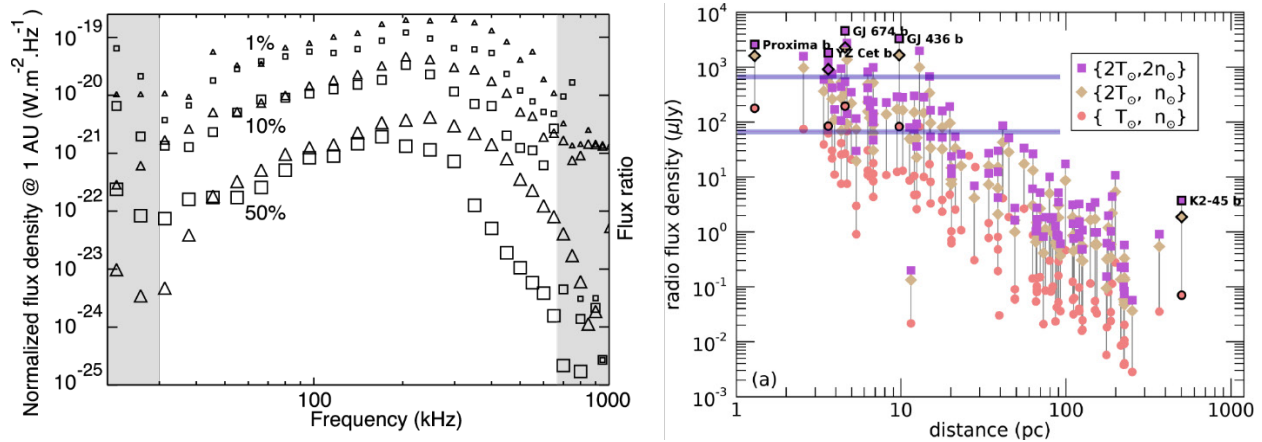

**Figure 3:** *Left:* The radio spectrum of Earth's auroral kilometric radiation for 50%, 10% and 1% of the time. The flux density across the entire detectable band can vary by factors of a few hundred [3]. *Right:* The predicted flux densities (at frequencies <280 kHz) for a sample of known terrestrial exoplanets orbiting M dwarfs [5]. The two horizontal lines highlight the sample of planets detectable by FARSIDE in a 1000-hour integration at 200 kHz (top) and in a one-hour integration where peak flux density is observed to increase by factors similar to observed for the Earth's AKR (bottom). A large sample of planets, including some nearby candidate habitable planets (such as Proxima b), are detectable.

## 1.2 The Dark Ages Redshifted 21-cm Global Spectrum

FARSIDE enables a precision measurement of the Dark Ages 21-cm all-sky spectrum at frequencies 10-40 MHz corresponding to redshifts z=130-35. Such observations have enormous potential as a powerful new test of the standard ΛCDM cosmological model in the early Universe and would provide constraints on any exotic physics of dark matter [24]. Figure 4 shows models and observations of the global spectrum in the Dark Ages and Cosmic Dawn.

Recent results [25, 26] from the *Experiment to Detect the Global Epoch of Reionization (EoR) Signature* (EDGES) suggest the presence of a strong feature in the 21-cm spectrum at ≈78 MHz (z≈17), within the range expected for the "Cosmic Dawn trough." Figure 4 shows deviations from the standard cosmology for phenomenological models of added cooling to the primordial neutral hydrogen [27] that could be produced by e.g. previously unanticipated interactions between baryons and dark matter particles.. The black dashed curve corresponds to the standard ΛCDM cosmology model containing Pop III stars [28, 29]. While the difference in the observed EDGES redshift can be explained within the standard model, the drop in brightness is about 3 times greater than expected. The "Dark Ages trough", at $\nu < 30$ MHz and inaccessible from the ground, is produced purely by cosmology and thus cleaner because there are no stars to complicate the signal. This trough is a prime target for FARSIDE.

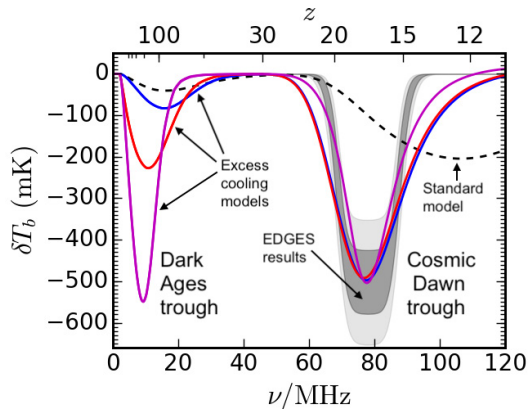

**Figure 4. The Dark Ages 21-cm absorption trough is a sensitive probe of cosmology.** The black dashed curve shows the brightness temperature (relative to the radio background) in the standard cosmological model with adiabatic cooling. The shape at z>30 is independent of astrophysical sources. The grey contours show the 1 & 2-sigma absorption bands inferred by EDGES. The solid curves are parametric models that invoke extra cooling to match the amplitude in the EDGES signal but also dramatically affect the dark ages absorption trough at z>50.

If the measured redshifted 21-cm signal differs from that of the standard cosmological model, new physics is required.

The major challenge is the presence of bright foregrounds that are >$10^4$ times that of the 21-cm signal, which together with chromaticity of the antenna beam, can introduce added frequency structure into the observed foreground brightness temperature spectrum. FARSIDE mitigates these effects in several important ways. First, the array would provide high resolution maps of the sky foreground at multiple relevant frequencies which facilitates the extraction of the 21-cm signal using a data analysis pipeline that we have built based on pattern recognition algorithms informed by training sets [30] that can be obtained from sky foreground observations, lab instrument measurements, beam simulations and 21-cm signal theory. Second an orbiting calibration beacon would permit us to map the antenna beam in the far-field for the first time enabling correction for beam chromaticity effects. Approximately $5000/\sqrt{N}$ hours of integration time, where N is the number of antenna elements used to measure the global signal, would separate the standard cosmology from added cooling models at >$5\sigma$.

## 1.3　Additional Science

*Heliophysics:* FARSIDE offers the possibility to detect and image solar radio bursts out to large heliocentric distances. Resolved imaging of is essential for establishing the location and spatial structure of the emission and thereby the physics behind the emission [2].

*Outer Solar System Planets:* FARSIDE would detect and monitor the auroral and lightning-generated radio emission from the outer gas giants, including Uranus and Neptune In fact, FARSIDE would offer the unique possibility of searching for radio emission from large bodies beyond Neptune out to 100s of AU. This includes, but is not limited to, the putative Planet 9 [31]. The expected flux density of Planet 9 is ~10 mJy below 1 MHz, if a planetary magnetosphere is present.

*Sounding of the Lunar Sub-surface:* FARSIDE has the potential to sound the mega-regolith and its transition to bedrock expected at ~2 km below the surface [32]. The Lunar Radar Sounder (LRS) onboard the KAGUYA (SELENE) spacecraft has provided sounding observations of the lunar highlands [33] and found potential scatterers in hundreds of meters below the subsurface. However, the results are inconclusive due to surface roughness. FARSIDE, by virtue of being on the surface, would not be affected by roughness. Data from a calibration beacon in orbit could be synthesized to identify deep scatterers and the transition to bedrock at km depths by virtue of the low frequencies, which are significantly more penetrating. Deep subsurface sounding can also be performed passively using Jovian bursts from 100 kHz to 20 MHz [34]. The array covers a 10 km × 10 km area on the lunar highlands which could provide a three dimensional image of the highland subsurface structure.

*Tomography of the Local ISM:* The full three-dimensional structure of the warm interstellar medium (WIM) in the solar neighborhood can be inferred through mapping the emissivity as a function of frequency over a wide range of frequencies with FARSIDE.

## 2 TECHNICAL OVERVIEW

**Table 1:** Baseline FARSIDE Specifications.

| Quantity | Value |
| --- | --- |
| Antennas | 128 × 20 m length dual-polarization dipoles |
| Frequency Coverage | 200 kHz – 40 MHz (1400 × 28.5 kHz channels) |
| Field of View (FWHM) | > 10,000 deg$^2$ |
| Spatial Resolution | 10 degrees @ 200 kHz / 10 arcminutes @ 10 MHz |
| System Temperature[a] | 9 × 10$^5$ K @ 200 kHz / 2.7 × 10$^4$ K @ 10 MHz |
| Effective Collecting Area[b] | ~ 18.5 km$^2$ @ 200 kHz / 6700 m$^2$ @ 10 MHz |
| System Equivalent Flux Density (SEFD) | 128 Jy @ 200 kHz / 1.1 × 10$^4$ Jy @ 10 MHz |
| 1σ Sensitivity (60 seconds; bandwidth = ν/2) | 40 mJy @ 200 kHz[c] / 0.5 Jy @ 10 MHz |
| 1σ Sensitivity (1 hour; bandwidth = ν/2) | 5 mJy @ 200 kHz[c] / 60 mJy @ 10 MHz |
| 1σ Sensitivity (1000 hours; bandwidth = ν/2) | 150 μJy[d] @ 200 kHz[c] / 2 mJy @ 10 MHz |

[a] System temperature includes contribution from the sky and ground due the absence of a ground screen.
[b] Effective area is impacted by loss of gain into the ground due to absence of a ground screen.
[c] Sensitivity calculations at 200 kHz assume night time conditions.
[d] Deep confusion-free integrations are possible < 3 MHz due to the absence of extragalactic sources.

### 2.1 Unique Environment of the Lunar Farside

Ground-based radio astronomy is limited to >10 MHz, due to attenuation in the ionosphere. Moving above the atmosphere opens an additional 2 orders of magnitude in frequency range below which observations become limited by the plasma density of the solar wind at Earth orbit. However, the sensitivity of an Earth-orbiting radio telescope is increasingly limited below ~1 MHz. This is partially due to the presence of anthropogenic radio frequency interference (RFI) and the intense auroral kilometric radiation (AKR) produced by Earth. More fundamentally, the system noise below 1 MHz in Earth orbit is limited by the electrons in the solar wind colliding with the antenna and inducing currents [35]. Together, these factors prevent sensitive searches for exoplanet radio emission below 1 MHz. By contrast, the lunar farside provides a unique environment for radio astronomy within the inner solar system, with > 100 dB attenuation of both RFI and AKR. Most importantly, a plasma cavity exists on the surface, particularly on the night time side, that provides sufficient isolation for Galactic noise dominated observations down to ~200 kHz. Within this environment, unique in the inner solar system, the effective collecting area of a dipole becomes very large reaching ~km$^2$ at 200 kHz, while the system temperature remains largely unchanged (optically thick Galactic synchrotron).

### 2.2 Proposed Architecture and Array Configuration

FARSIDE would consist of a central Base Station and 128 antenna nodes distributed across a 10 km diameter area (Figure 5). The nodes are distributed along 8 independent tethers, which are deployed to form 4 petals with a deliberate asymmetry to improve imaging performance. For each petal, the deployment rover would carry a set of antenna nodes out from the base station, unreeling the tether, before returning to base along a different path, continuing to unreel the tether and attached nodes. The Base Station would provide power, signal processing, and telecommunications back to Earth (via a relay). The individual antenna nodes are a crossed dipole with preamplifier, sending the two signals back to the Base Station via optical fibers as an analog signal. The antenna nodes are powered in series similar to the scheme used for transoceanic cables, so only two power conductors are needed (although 6 conductors are used for redundancy).

### 2.3 Antenna and Receiver

Each Antenna Node consists of a small (5 × 5 × 10 cm) aluminum housing containing two receiver channels, each comprised of a high input impedance low noise preamplifier using the flight proven [36] OPA656 operational amplifier driving a laser diode feeding an optical fiber back to the Base

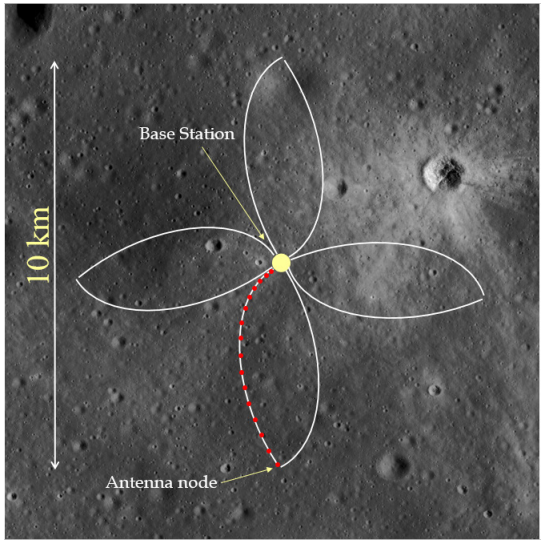

**Figure 5:** FARSIDE will consist of 128 antenna nodes deployed by a rover from a central base station, arranged in a petal configuration.

Station. The input to the amplifier is a 20-meter-long dipole antenna that is two wires, each 10 meters long. One of the dipoles is embedded in the tether and the other is laid out roughly at right angles to the tether by two STACER deployable elements. To save tether mass, all the nodes in a "petal" are connected in series along the tether so there are only 2 power wires in the tether. A ~300V voltage power source at the base station injects a constant current, and a switching converter at each node produces the voltages needed by the node.

A Radioisotope Heating Unit (RHU) is baselined to provide heat to maintain the temperature of the electronics during the long lunar night. A thermal switch disconnects the radiator from the electronics when the temperature drops. An alternate approach to the RHU is to package the electronics so that it can survive the -200 C night. The tether that connects the 16 nodes to the base station is a thin (0.025 mm) polyimide (Kapton) tape 1.5cm wide with 32 x 0.125 mm diameter optical fibers and 6 AWG 30 (0.25 mm) aluminum wires. The wires are arranged in groups of 3 for redundancy, so that if a single wire breaks, the remaining 2 can carry the current. The tether for each petal is wound on a spool that is carried by the deployment rover.

The receiver does not use a resonant antenna, rather it is a voltage probe that measures the electric field and uses a high impedance amplifier to measure the open circuit voltage of a dipole antenna. The combination of the antenna's impedance and the input impedance of the amplifier form a voltage divider. The resistive component of the impedance is composed of the radiation resistance and the loss resistance. The radiation resistance is $20\pi^2(L/\lambda)^2$, the loss resistance is due to dielectric losses and finite conductivity of the elements, and was modeled using the NEC4.2 numerical model [37] assuming $\varepsilon_r = 3+0.005j$, from [38]. The effective collecting area and sky-noise estimated from these calculations (e.g. Figure 7) are the basis for the sensitivity estimates presented in Table 1.

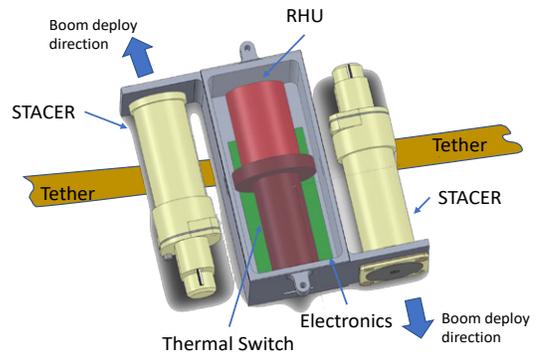

**Figure 6:** Mechanical design of the FARSIDE antenna node.

The electrical performance of a wire antenna near, on, or below, a dielectric surface has been studied in the context of ground penetrating radar, and for various microwave detectors. Considering a transmitting antenna, it radiates more power into the dielectric than the air (or space) by $\varepsilon_r^{3/2}$ [39], a power ratio of about 5 (7dB) for typical regolith $\varepsilon_r = 3$. In addition, the reflected wave from the regolith destructively combines with the direct wave reducing the effective height of the antenna to 73% of the free space effective height. FARSIDE uses 10-meter dipole arms, producing an effective height of approximately 7 meters. Since the antenna is so close to the regolith, in terms of wavelength, the use of a ground screen (as used at higher frequencies in observatories such as LWA and LOFAR) is impractical.

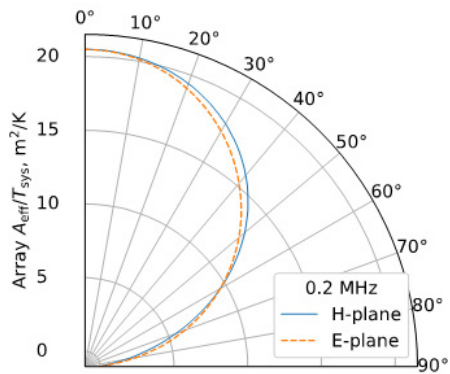

**Figure 7:** Effective Area/System Temperature as a function of zenith angle at 200 kHz.

### 2.4    Lander and Base Station

The 590 kg FARSIDE base station would use a commercial lander, assumed for the purposes of this study to be similar to the Blue Origin Blue Moon lander. The electronics are fully dual string. The Base Station houses the FX correlator - each of the eight spokes feeds an F-engine board which receives the 32 optical signals (two polarizations per antenna node) and performs the frequency channelization in space qualified FPGAs(Field Programmable Gate Arrays). The output of those boards are corner-turned to X-engine boards that perform full cross-correlation of all antennas, also in FPGAs. The integrated output of the correlators is passed to the JPL-developed, radiation-tolerant, FPGA-based, Miniaturized Sphinx onboard computer and stored for relay to Earth. 192 GB of storage is provided, about 3 days at full rate. The Sphinx is also responsible for system management, engineering housekeeping telemetry, and control functions.

With no direct path to Earth, the Base Station communicates with the Lunar Gateway (or an alternative orbiting relay) using a conventional 8 Mbps RF link through a 90 cm gimbaled high gain antenna (HGA) during the 85% of the time that the Gateway is in view. Low gain antennas provide safe mode communications or when the HGA is unusable (i.e. if the position of the Gateway is not known). The baseline data rate for the array is ~6 Mbps, accommodated by the RF link. The telecom subsystem is essentially identical between the rover and the base station - both need to communicate to the gateway, and a relay would not work since the rover would travel farther away than the 1-2 km visible horizon. The baseline concept uses a pair of eMMRTGs (enhanced Multi Mission Radioisotope Thermoelectric Generator) with EOM (end of mission) power of 271W providing 30% contingency over normal requirements. Batteries provide additional reserve capacity to accommodate peak loads (311W for Science at night). An active pumped fluid loop manages the waste heat. The power system is conventional with the addition of a boost converter to take the Base Station bus voltage and boost it to the higher voltage (~300V) needed to feed the series strings of 16 antenna nodes.

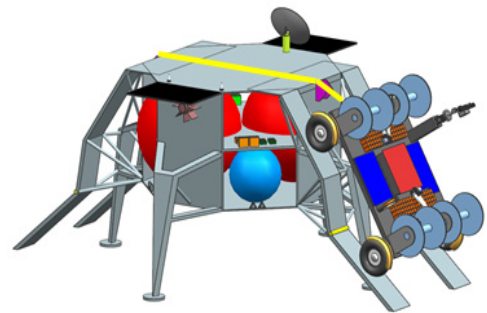

**Figure 8:** Conceptual design of the FARSIDE rover being deployed from the Base Station.

### 2.5    Deployment Rover

The 420 kg rover is based on the Apollo Lunar Roving Vehicle and includes 1.6 m$^2$ of solar array and a 30 Ah battery to stay alive during the lunar night. 4 RHUs are baselined to keep the avionics warm at night, and during the day, a 1.3 m$^2$ radiator would reject the heat during operations. The rover carries the same flight computer and telecom package as the base station. After landing, the rover would be deployed down a ramp to the lunar surface, using a winch during the initial steep descent. Ramps are provided on both sides for redundancy, in the event that the landing site makes it difficult to descend on one side or the other.

The rover includes 4 hazard avoidance cameras and 2 navigation cameras on a mast which are used by human operators on Earth. Every 5 meters (about 3-4 minutes at the expected driving speed) a set of images would be collected and sent back to Earth, the teleoperators would confirm the next 5 meter driving increment, and the process repeats, roughly every 7 minutes. An antenna node is deployed after every 3-4 hours of driving. Allowing 1 hour to deploy each antenna node, the trip out from the base to deploy 16 nodes would take 80 hours, and another 80 hours to return to base, deploying the second spoke. We have allowed one lunar day (14 Earth days) to completely deploy one petal (two spokes) to allow for intervals when the Lunar Gateway is not in view, and for contingencies to route around obstacles. During the lunar night, the rover goes into deep sleep mode. After 4 months, all petals would have been deployed.

## 2.6     Scientific Data Products

FARSIDE would collect full cross-correlation data every 60 seconds, in 1400 channels of 28.5 kHz width each, for a total data rate of 65 GB per 24-hour period. All visibility data would be transferred via the Lunar Gateway (or orbiting relay) to Earth. A snapshot image of the ~10,000 deg$^2$ area within the half-power primary beam of the antennas would be produced for each of the 1400 channels in each of Stokes IQUV, as well as a combined full-band image. Dynamic spectra at the location of every stellar/planetary system within 10 pc would be produced from these data to search for Type II/III radio bursts and enhanced planetary auroral emission. In addition, data from each lunar day would be combined via the technique of m-mode analysis [40] to produce deep all-sky images that would precisely map the synchrotron and absorbed free-free emission in our Galaxy, as well as providing the best means for deep searches for the median flux densities of exoplanets. The use of snapshot all-sky imaging together with deep integration all sky imaging via m-mode analysis has already been demonstrated for the OVRO-LWA, albeit at >100 times higher frequencies than accessible to FARSIDE [41, 42]. Simultaneously auto-correlation power spectra for each antenna would be collected every 60 seconds, which together with the m-mode foreground maps, would deliver the radio spectrum of the Dark Ages 21-cm signature.

While calibration of the array can be carried out using astronomical sources, both the bandpass of each antenna and the direction independent and direction dependent (antenna beams) gain parameters for the array would be primarily determined via an orbiting calibration beacon.

## 3     TECHNOLOGY DRIVERS

The FARSIDE array concept uses flight-proven hardware, with limited required technology development that is well understood. Notably, the rover would cover ~45 km in deployment of the array, which would exceed any previous lunar rover (Apollo 17 covered ~36 km) and would equal the record of the Opportunity rover. However, this record would represent an incremental improvement, rather than a transformative one. The procurement of >100 RHUs does represent a significant logistical concern as this level of quantity has not been produced previously.

## 4     ORGANIZATION, PARTNERSHIPS, AND CURRENT STATUS

FARSIDE is a collaboration between the University of Colorado Boulder (PI Jack Burns), the California Institute of Technology (Co-PI and Project Scientist Gregg Hallinan), Arizona State University (Co-I Judd Bowman), NASA Goddard Space Flight Center (Co-I Robert MacDowall), the National Radio Astronomy Observatory (Co-I Richard Bradley), and the Jet Propulsion Laboratory (management and lead instrument organization).

FARSIDE was initially developed, and funded, as part of the research program for the Network for Exploration and Space Science (NESS, https://www.colorado.edu/ness/), a NASA Solar System Exploration Research Virtual Institute (SSERVI). Paul Hertz then selected FARSIDE for a Probe-class design study, recognizing the programmatic synergy between Decadal-level science and NASA's

Artemis program of human/robotic exploration of the Moon. The FARSIDE team selected JPL as its partner for the design study.

JPL employed its standard concept development process for designing the mission, comparing projected performance to science requirements, assessing technical feasibility, and estimating cost. A preliminary Science Traceability Matrix (STM) was developed to sharpen science objectives and provide requirements traceability for prospective point designs. JPL's concurrent design facility, Team X, performed a preliminary study to consider candidate architectures and selected one that met performance requirements, was technically feasible within the ∼3000 kg mass constraint of a large lunar lander currently under design by NASA, and satisfied the $1B cost constraint for Probe-class missions. The architecture from that study formed the basis for a more in-depth study by Team X, which conducted an instrument study for the receiver node, and mission studies for the base station and deployment rover. The final result was the more detailed point concept design described in this paper, which is pre-decisional and provided for planning and discussion purposes only.

## 5 SCHEDULE

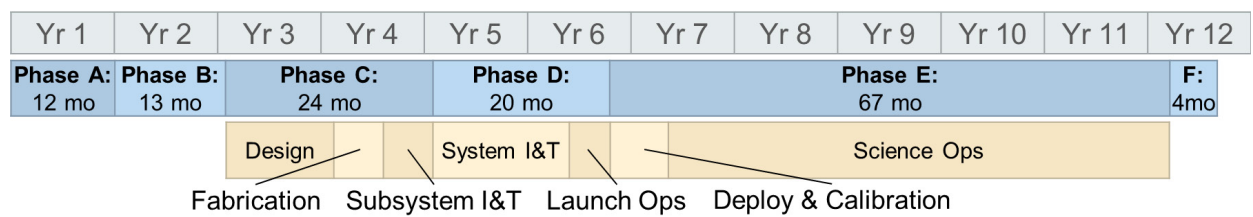

## 6 COST ESTIMATES

JPL's Integrated Design Facility (Team X) explored three architectures to initially assess technical feasibility and cost[2]. The $1B Probe-class mission nominal cost constraint was required for a successful candidate design. Table 2 shows the resulting total mission cost, including Operations Mission Phases E-F, and the cost of the MMRTG and RHU elements. It is roughly consistent with the cost constraint within current uncertainties.

Costs were estimated using the standard NASA Work Breakdown Structure (WBS). The basis of estimate for costing the eMMRTGs and RHUs is the guidance provided in the Discovery 2019 Announcement of Opportunity. Receiver node cost for the Theoretical First Unit (TFU) was derived from the NASA Instrument Cost Model (NICM), based on mass and power; the cost for additional receivers followed the Wright Learning Curve model [43]. The deployment rover cost was scaled from MER based on an estimate of the rover mass. The base station cost used the standard Team X cost modeling process, based on preliminary design requirements. The other WBS elements were costed as percentages of these costs based on historical actual costs for NASA flight missions.

Table 2: Cost summary for FARSIDE[1]

| Cost Summary (FY2019$M) | Team X Estimate | | |
|---|---|---|---|
| | CBE | Res. | Cost + Reserve |
| Total Cost | $1080M | 27% | $1330M |
| MMRTG + RHU | $70M | 0% | $70M |
| Launch Vehicle | $150M | 0% | $150M |
| Development & Ops Cost | $865M | 29% | |
| Development Cost | $800M | 30% | $1040M |
| Phase A | $8M | 30% | $10M |
| Phase B | $70M | 30% | $90M |
| Phase C/D | $720M | 30% | $940M |
| Operations Cost (Phase E/F) | $65M | 15% | $75M |

---

[2] The cost information contained in this document is of a budgetary and planning nature and is intended for informational purposes only. It does not constitute a commitment on the part of JPL and/or Caltech.